\documentclass[a4paper,fleqn,12pt]{cas-sc}
\usepackage[numbers]{natbib}
\usepackage{xcolor}
\ExplSyntaxOn
\cs_set:Npn \__first_footerline: {}
\ExplSyntaxOff
\begin{document}

\shorttitle{Spectroscopy of {Ho$^{3+}$ doped Y$_{2}$SiO$_{5}$}}
\shortauthors{Mothkuri et~al.}
\title [mode = title]{Laser Site-Selective Spectroscopy and Magnetic Hyperfine Splittings of {Ho$^{3+}$ doped Y$_{2}$SiO$_{5}$}} 
\author[1,2]{Sagar Mothkuri}[]
\credit{Experiment, Methodology, Software, Analysis}

\affiliation[1]{organization={School of Physical and Chemical Sciences,University of Canterbury},
                addressline={PB4800}, 
                city={Christchurch},
                postcode={8140}, 
                country={New Zealand}}
\affiliation[2]{organization={Dodd-Walls Centre for Photonic and Quantum Technologies},
                country={New Zealand}}
\affiliation[3]{organization={Chimie ParisTech, PSL University, CNRS, Institut de Recherche de Chimie
Paris},
country={France}}
\affiliation[4]{organization={Thales Research and Technology, 1 Avenue Augustin Fresnel, 91767, Palaiseau},
country={France}}
\affiliation[5]{organization={Facult{\'e} des Sciences et Ing{\'e}nierie,  Sorbonne Universit{\'e}},
country={France}}
\author[1,2]{Michael F. Reid}[orcid=0000-0002-2984-9951]
\cormark[1]
\credit{data analysis and interpretation}
\ead{mike.reid@canterbury.ac.nz}
\author[1,2]{Jon-Paul R. Wells}[orcid=0000-0002-8421-6604]
\cormark[1]
\credit{Conceptualisation of this study, data analysis and interpretation}
\ead{jon-paul.wells@canterbury.ac.nz}
\cortext[cor1]{Corresponding authors}
\author[3,4]{Elo{\"i}se Lafitte-Houssat}
\credit{Crystal growth}
\author[3,5]{Alban Ferrier}
\credit{Crystal growth}
\author[3]{Philippe Goldner}
\credit{Crystal growth and advice on interpretation}

\begin{abstract}
  Laser site-selective spectroscopy and high-resolution absorption measurements have been used to determine 51 crystal-field energy levels for one of the Ho$^{3+}$ centres in Y$_{2}$SiO$_{5}$. This centre is denoted as Site 2 and has been tentatively assigned as the seven-fold coordinated centre.
  High resolution absorption measurements reveal complex hyperfine patterns that obey and approximate selection rule. The application of a magnetic field along the three optical axes reveals the presence of avoided crossings below 0.5 Tesla, in both the ground and excited states.
%{\color{red}[Sentence deleted.]}
%Above 0.5 Tesla, a breakdown in the selection rule is observed.

%
%~\\ \noindent Mike Version 06, 08/05/24: \today
\end{abstract}

%\begin{graphicalabstract}
%  % \includegraphics{figs/grabs.pdf}
% \includegraphics[width=0.5\textwidth]{graphical_abstract}
%\end{graphicalabstract}

% \begin{highlights}
% \item Identification of energy levels of Ho$^{3+}$ in  Y$_2$SiO$_5$
% \item Observation and modelling of magnetic hyperfine structure in absorption
% \item Approximate selection rules for hyperfine transitions in low symmetry sites
% \end{highlights}

%\printcredits


\maketitle

\section{Introduction} \label{sec:Intro}

Lanthanide-doped Y$_2$SiO$_5$  has attracted considerable interest as an option for the development of quantum-information technologies. The magnetic moment of yttrium is small and the common isotopes of Si and O have zero nuclear spin.
Consequently, decoherence due to nuclear spin flips is minimised and coherence times exceeding 1 minute for Pr$^{3+}$:Y$_2$SiO$_5$ and 6 h in the case of Eu$^{3+}$:Y$_2$SiO$_5$ ~\cite{Georg2013,Zhong2015} have been reported. Single-photon control of multiple ions has also been demonstrated  \cite{thompson2018}. 
The long coherence times are achieved using the zero-first-order Zeeman (ZEFOZ) technique, which exploits avoided crossings of hyperfine sub-levels to minimise sensitivity to magnetic field fluctuations. The calculations required to locate 
ZEFOZ points have used spin Hamiltonians ~\cite{Longdell2002}. However, spin-Hamiltonian calculations break down at high magnetic fields, due to mixing of electronic states. Calculation at the high magnetic fields used to obtain long coherence at telecom frequencies \cite{rancic2018}  are not possible with spin Hamiltonians. 

Crystal-field calculations ~\cite{carnall_systematic_1989,GoBi96,NeNg00,liu_electronic_2006} do not suffer from this problem but are challenging in low-symmetry sites since up to 27 crystal-field parameters are required.  We have recently developed computational methods that allow us fit a crystal-field Hamiltonian to not only electronic levels but also magnetic splittings and magnetic-hyperfine splittings in low-symmetry sites. These methods have been applied to Er$^{3+}$  ~\cite{Horvath2019,Jobbitt2021a} and other Kramers ions,  Ce$^{3+}$ \cite{Alizadeh2021}, Nd$^{3+}$ \cite{Alizadeh2023}, Sm$^{3+}$ \cite{Jobbitt_2022},  and Yb$^{3+}$ \cite{Zhou2020} doped Y$_2$SiO$_5$. 

Fits for non-Kramers ions in low symmetries are more challenging as the electronic states are non-degenerate. Consequently, magnetic-splitting data is more difficult to obtain. However, Ho$^{3+}$ has high electron-nuclear coupling and often has pseudo doublets that split under a magnetic field. A preliminary report ~\cite{SagarMothkuri2021} presented hyperfine structure in   transitions to the lowest levels of the $^{5}$I$_{7}$ multiplet for Ho$^{3+}$:Y$_2$SiO$_5$. In this work, laser site selective fluorescence measurements are used to construct a partial energy level scheme for the Ho$^{3+}$ site that we label site 2. Since we only have magnetic splitting information along three axes, there is insufficient data for a full crystal-field fit. However, wavefunctions can be determined from a calculation using well established Er$^{3+}$ crystal-field parameters \cite{Jobbitt2021a}. These wavefunctions are used to interpret complex Zeeman-hyperfine patterns observed for the $^{5}$I$_{7}$ and $^{5}$I$_{6}$ multiplets in high resolution absorption experiments using a four Tesla magnet.

\section{Experimental} \label{sec:Exp}

A Ho$^{3+}$:Y$_{2}$SiO$_{5}$  crystal with a Ho$^{3+}$ concentration of ${200}$\,ppm was prepared in the X2 phase  using the Czochralski method. The crystal was oriented using Laue backscattering. Y$_2$SiO$_5$ has
three perpendicular optical-extinction axes: the crystallographic $b$
axis and two mutually perpendicular axes labelled $D_1$ and $D_2$. In
our calculations we follow the convention of identifying these as the
$z$, $x$, and $y$ axes respectively \cite{sun_magnetic_2008}. The sample was a cuboid with the $D_1$ and $D_2$ and $b$ axes through the faces with dimensions of ${6.27\pm0.1}$ mm, ${6.72\pm0.1}$ mm, and ${4.85\pm0.1}$ mm respectively. 
The X2 phase of Y$_2$SiO$_5$ is a monoclinic crystal having C$^6_{2h}$ space group symmetry. The lattice constants are $a$ = 10.41 {\AA}, $b$ = 6.72 {\AA}, $c$ = 12.49 {\AA}, and $\alpha$ = 90$^{\circ}$, $\beta$ = 102$^{\circ}$39', $\gamma$ = 90$^{\circ}$. 
The yttrium ions occupy two crystallographically distinct sites, each with C$_1$ point-group symmetry. We follow the labelling convention used in Er$^{3+}$ ~\cite{Horvath2019,sun_magnetic_2008} and in Ref.~\  \cite{SagarMothkuri2021,Jobbitt2019} of referring to these two substitutional sites as site 1 and site 2, tentatively identified as six- and seven-coordinate.   

Infrared absorption spectra were obtained using a Bruker
Vertex 80 FTIR, with a maximum apodized resolution of 0.075\,cm$^{-1}$.
The entire optical path of FTIR was purged by N$_2$ gas. The sample was placed in a Janis CCS-150 closed-cycle cryostat and cooled down to a temperature of 10~K to perform absorption spectroscopy or alternatively screwed into the solenoid of a four Tesla magnet which was itself immersed in liquid helium. The magnet was powered by a 60 A current supply.
%{\color{red}
The measurements used unpolarized light and the propagation direction was along the magnetic field axis.
%}
Laser site-selective spectroscopy was performed using a PTI  GL-302 tunable dye laser pumped by a pulsed nitrogen (N$_{2}$) laser, along with a Horiba iHR550 single monochromator, water-cooled Hamamatsu R2257 visible photomultiplier tube (PMT), and an air-cooled Hamamatsu H10330C infrared PMT. The signal was then fed to the SR250 boxcar averager before finally recording the data on LabVIEW software through a 3GHz Tektronix DPO7104 digital oscilloscope. Again the sample was placed in a Janis CCS-150 closed-cycle cryostat having a base temperature of 10~K. A Lakeshore 325 temperature controller was used to adjust the sample temperature via a resistive heater, attached to the back of the sample cold finger.

\section{Theoretical background}
\label{sec:Theory}
The Hamiltonian for the $4f^N$ configuration may be written as
\cite{carnall_systematic_1989,GoBi96,liu_electronic_2006}

\begin{equation}
  H = H_{\mathrm{FI}}
  + H_{\mathrm{CF}}
  + H_{\mathrm{Z}}
  + H_{\mathrm{HF}}
  + H_{\mathrm{Q}} ,
  \label{eq:crystalfieldhamiltonian}
\end{equation}
where
$H_{\mathrm{FI}}$ is the free-ion contribution,
$H_{\mathrm{CF}}$  the crystal field interaction,
$H_{\mathrm{HF}}$ the nuclear magnetic dipole hyperfine interaction,
$H_{\mathrm{Z}}$ the Zeeman interaction and
$H_{\mathrm{Q}}$ the nuclear quadrupole interaction. 

The free-ion Hamiltonian  may be written as
\begin{align}
 H_{\mathrm{FI}} &= E_\text{avg} + \sum_{k=2,4,6} F^k f_k + \zeta   
   \sum_i \mathbf{s}_i \cdot \mathbf{l}_i
 \nonumber\\
 &+\alpha L(L+1) + \beta G(G_2) + \gamma G(R_7) 
                 + \sum_{k = 2,3,4,6,7,8} T^k t_k
 \nonumber \\  
 &  + \sum_{k=0,2,4} M^k m_k + \sum_{k=2,4,6} P^k p_k , 
  \label{eqn:h_fi_defn}
\end{align}
where
$E_\text{avg}$ is a constant configurational shift, $F^k$ are the Slater
parameters characterizing aspherical electrostatic repulsion, and
$\zeta$ is the spin-orbit coupling constant. The sum in the spin-orbit term is over the $4f$ electrons. The other terms 
parameterize two- and three-body interactions, as well as higher-order
spin-dependent effects
\cite{carnall_systematic_1989,liu_electronic_2006}.

The crystal-field Hamiltonian has the form 
\begin{equation}
    H_{\mathrm{CF}} = \sum_{k,q} B^{k}_q C^{(k)}_q , 
\end{equation}
with k = 2, 4, 6, q = -k...k. The $B^{k}_q$ are crystal field parameters and the $C^{(k)}_q$ are spherical tensor operators. In $C_1$ symmetry, all non-axial ($q \neq 0$)\ $B^{k}_q$ parameters are complex, so there are 27 real parameters. In this low symmetry all electronic states are non-degenerate for  non-Kramers ions such as Ho$^{3+}$. 

The Zeeman interaction written as:
\begin{equation}
    %H_{\mathrm{Z}} = \mu_B\sum_{i}^{N} B (l_i +2s_i)
  H_{\mathrm{Z}} = \mu_B\mathbf{B} \cdot (\mathbf{L} +\mathbf{2S})
  %= \mu_B g\mathbf{B}\cdot \mathbf{J}, 
  \label{eqn:Zeeman}
\end{equation}
where B is the magnetic field. 

The electronic states of Ho$^{3+}$ are coupled to the nuclear spin through the hyperfine interaction. The nuclear spin is 7/2, so there are eight electronic-nuclear states for each crystal-field level. For holmium, the magnetic-hyperfine interaction is much larger than the nuclear-quadrupole interaction~\cite{Wells2004,Boldyrev2019}, so here we consider only the magnetic-hyperfine contribution. 
Within a $J$ multiplet the
%{\color{red}
  nuclear magnetic dipole hyperfine
%}
interaction may be written as 
\begin{equation}
    H_{\mathrm{HF}} = A_{J}\mathbf{J}\cdot \mathbf{I}  , 
  \label{eqn:AdotI}
\end{equation}
where $A_J$ is the magnetic hyperfine coupling constant of a multiplet, \textbf{J}, \textbf{I} are the total angular momentum and the nuclear spin operators respectively. 

%{\color{red}[Text deleted.]}

% In this  paper we use two approaches to the calculation of magnetic-hyperfine transitions. The first is a model calculation using a small basis set and the second is a full crystal-field calculation.

%\subsection{Model calculations}

%%%%%%%%%%%%%%%%%% matrix element table. 
\begin{table}[width=0.495\textwidth,cols=4,pos=tb]
\centering
\caption{\label{tab:Joperator_II}
Absolute values of matrix elements of the angular momentum operators from crystal-field calculations. The diagonal matrix elements are not shown but are of the order of 10$^{-4}$.}
 \begin{tabular*}{\tblwidth}{@{} LLLL@{} }
   \toprule
     Energy  & ${J}_{x}$ & ${J}_{y}$ & ${J}_{z}$ \\
     level & & &\\
     \midrule
    $|\langle Z_1|{J}_{i} |Z_2\rangle|$ & 3.01 & 2.75 & 5.12  \\
    \\
    $|\langle Y_1|{J}_{i} |Y_2\rangle|$ & 3.01 & 2.76 & 4.99  \\
    \\
    $|\langle A_1|{J}_{i} |A_2\rangle|$ & 2.51 & 1.99 & 3.78  \\
  \bottomrule
    \end{tabular*}
  \end{table}

The calculations followed the approach used in our previous work~\cite{SagarMothkuri2021}. 
Crystal-field parameters for Er$^{3+}$ in Y$_2$SiO$_5$ from Ref.~\cite{thesis:Horvath2016} and  free-ion parameters for Ho$^{3+}$ from Ref.~\cite{carnall_systematic_1989} were used to calculate electronic energy levels and their associated wavefunctions for the lowest two electronic states the $^5$I$_8$, $^5$I$_7$, and $^5$I$_6$ multiplets (Z, Y, and A in Dieke's notation \cite{carnall_systematic_1989}).
These wavefunctions were used to calculate matrix elements of the  angular momentum operators ${J}_{x}$, ${J}_{y}$ and ${J}_{z}$ for each pair of electronic states (Table~\ref{tab:Joperator_II}). Zeeman and hyperfine splittings of the 16 electronic-hyperfine states  $|JM_J,IM_I\rangle$ for each pair were calculated using these matrix elements. The spacings between the electronic states were treated as free parameters.  For simulated spectra, magnetic dipole moments for the transitions were calculated. These were weighted with relevant Boltzmann factors.
%{\color{red}
  A Lorentzian line shape with
%}
line width (FWHM) of 0.12 cm$^{-1}$ was used.

% \subsection{Crystal-field fits}
% \label{sec:CF}

%  A detailed description of the crystal-field analysis approach employed can be found in Ref. ~\cite{Horvath2019,thesis:Horvath2016}.  
% The initial parameters were the  free-ion parameters of Ho$^{3+}$ from Ref.~\cite{carnall_systematic_1989} and the crystal-field parameters of Er$^{3+}$:Y$_{2}$SiO$_{5}$ from Ref.~\cite{thesis:Jobbitt2021}. Hyperfine and quadrupole coupling constants of 0.048 and 0.06 respectively were estimated by comparing the calculations with experiment. Since the magnetic splitting data was limited to three axes, the fits cannot be considered definitive. However, they do give an indication of how the model calculations break down at high magnetic fields. 

\clearpage

\section{Results and Discussion}
\label{sec:R&D}

The energy levels for Site 2  inferred from a combination of laser site-selective fluorescence and absorption spectra  are summarized in Table ~\ref{tab:laser_excitation}.

\subsection{Absorption and laser site-selective fluorescence spectroscopy}\label{sec:Emission}

\begin{figure*}
\centering
\includegraphics[width=\textwidth]{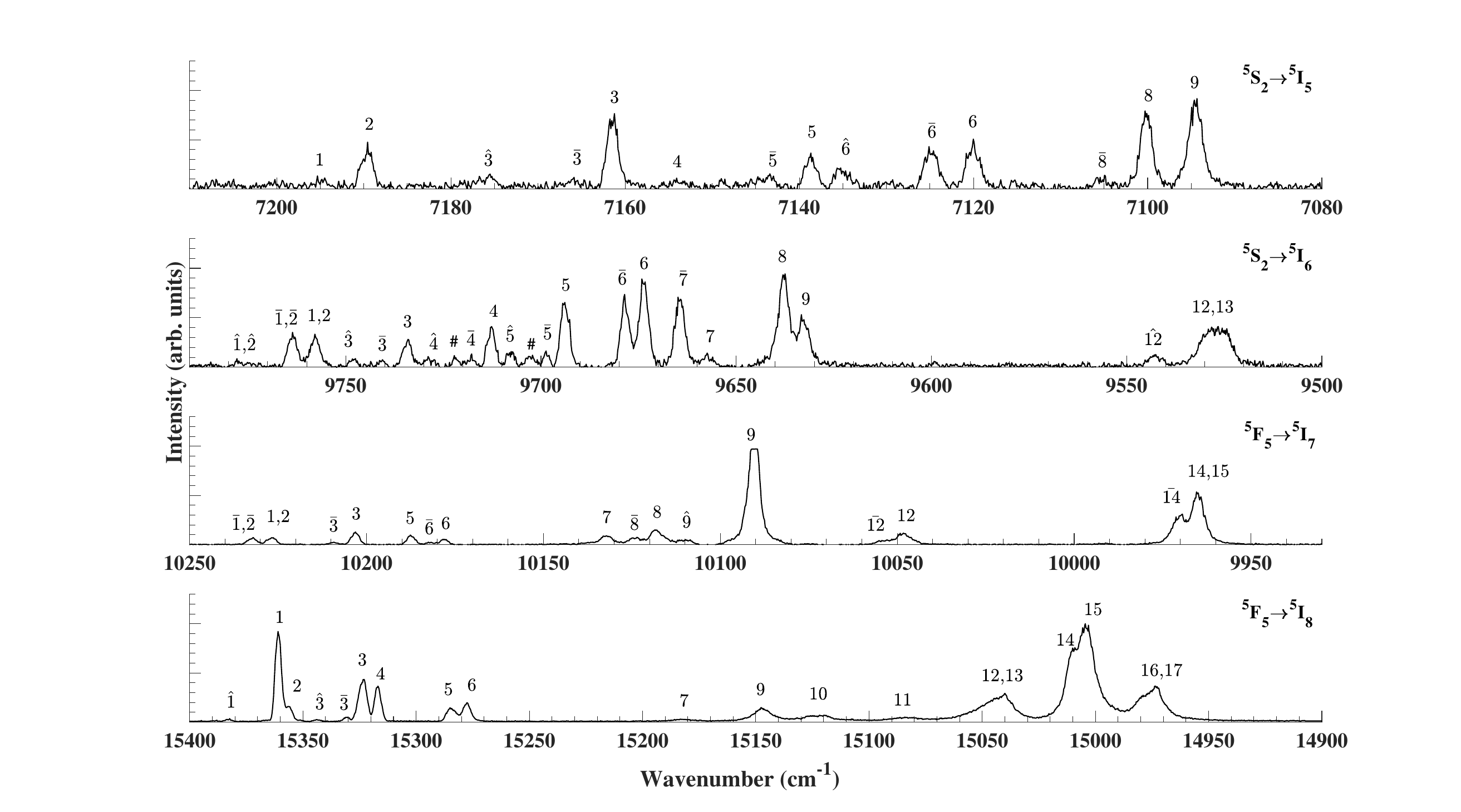}
\vspace{4mm}
\caption{\label{fig:emission} 10~K Site 2 laser site-selective
  fluorescence spectra of $^5$I$_8$(Z),$^5$I$_7$(Y),$^5$I$_6$(A), and $^5$I$_5$(B) in Ho$^{3+}$:Y$_{2}$SiO$_{5}$. The fluorescence spectra were obtained for excitation at a wavelength of 540.9 nm (18488 cm$^{-1}$). The peaks are labelled by their state level within the multiplet. Thermally excited transitions are also present and these are represented by
%{\color{red}[(symbol - deleted)]}
 `-' $\&$ ` $\hat{}$ ' indicating transitions from the first and second excited levels of the emitting multiplet. `$\#$' indicates an unassigned spectral feature. } 
\end{figure*}

Figure~\ref{fig:emission} shows laser site-selective fluorescence spectroscopy of the $^5$I$_8$ (Z), $^5$I$_7$ (Y), $^5$I$_6$ (A) and $^5$I$_5$ (B) multiplets, measured for a sample cooled to 10~K and excited using Coumarin 540A dye. The $^5$F$_5$$\rightarrow$$^5$I$_8$, $^5$F$_5$$\rightarrow$$^5$I$_7$, $^5$S$_2$$\rightarrow$$^5$I$_6$ and $^5$S$_2$$\rightarrow$$^5$I$_5$ fluorescence spectra for Site 2, were obtained by exciting the sample at a wavelength of 540.9 nm (18488 cm$^{-1}$). The  measured fluorescence was weak. This is attributed to the high phonon energies of the YSO host. The strongest visible fluorescence was observed from $^5$F$_5$ with fluorescence from $^5$S$_2$ strongly quenched by non-radiative relaxation. Transitions from thermally populated, excited state levels can be observed in the spectra. These transitions were used to provide extra assignments, such as the lowest three crystal-field levels of $^{5}$S$_{2}$.

The fluorescence from $^5$F$_5$ to $^5$I$_8$ (Figure ~\ref{fig:emission}) indicates a ground state energy level separation Z$_1$-Z$_2$ of 4.9 ($\sim$5) cm$^{-1}$. This is consistent with the measured absorption spectra for Site 2. A comparable separation of 5 cm$^{-1}$ is measured for the $^{5}$F$_{5}$ D$_1$-D$_2$ levels. Such pseudo-doublets, comprised of close lying singlets states, are common in Ho$^{3+}$ and yield a rich and complex hyperfine structure. The absorption spectra for $^5$I$_7$, $^5$I$_6$, $^5$I$_5$ indicate similarly close splittings for the lowest lying levels of these mulitplets although this is not observable in lower resolution fluorescence spectra.

\begin{figure*}
%\begin{sidewaysfigure*}[h]
\centering
\includegraphics[width=\textwidth]{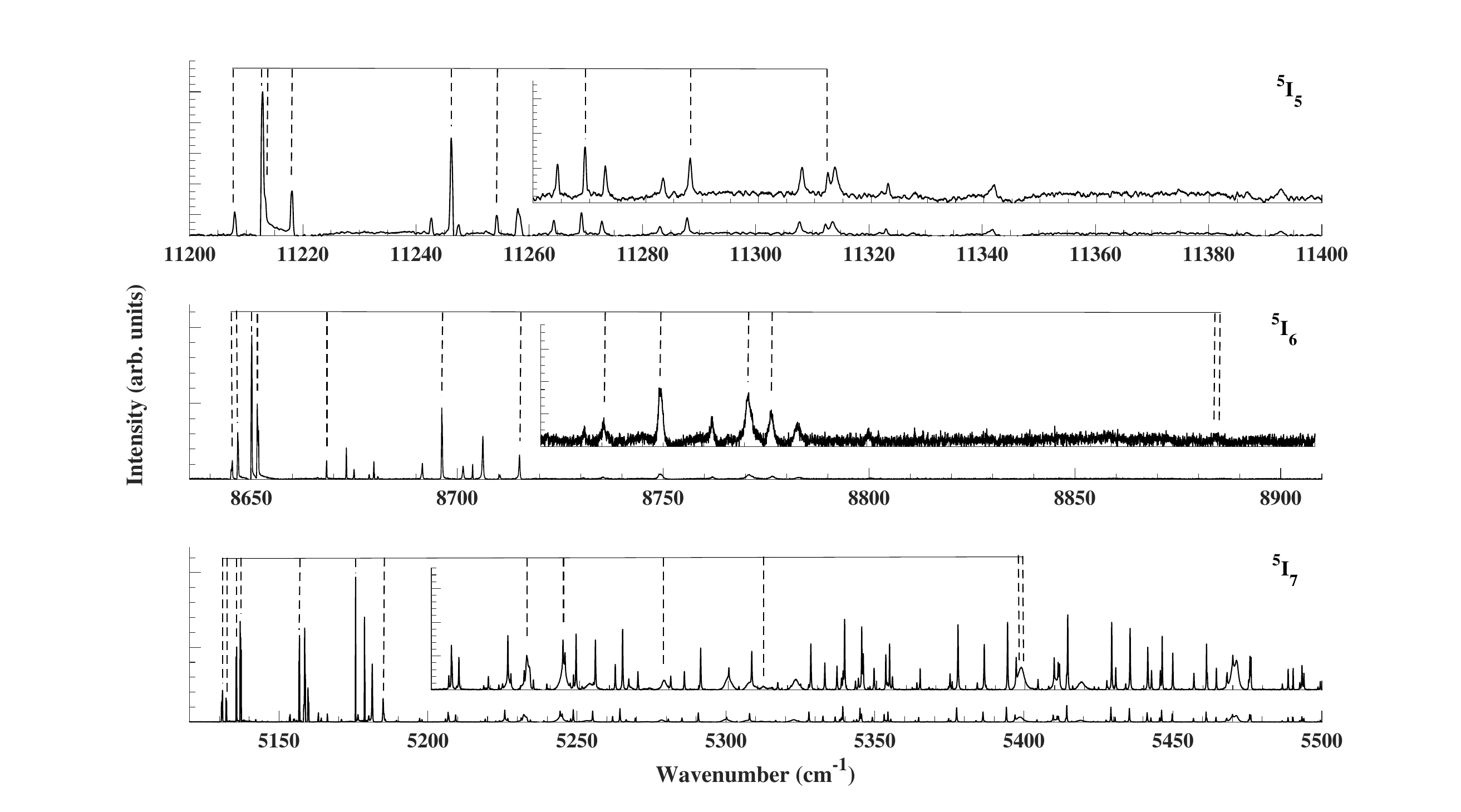}
\vspace{4mm}
\caption{\label{fig:absorption} 4.2~K absorption spectra of 
 $^5$I$_7$(Y),$^5$I$_6$(A), and $^5$I$_5$(B) in Ho$^{3+}$:Y$_{2}$SiO$_{5}$. Site 2 peaks are labelled by black dotted lines. The site 1 peaks are not marked for simplicity and the remaining (very sharp features) are due to atmospheric water absorption.} 
\end{figure*}

Figure~\ref{fig:absorption} shows 4.2~K absorption spectra for the $^5$I$_7$ (Y), $^5$I$_6$ (A) and $^5$I$_5$ (B) multiplets. Transitions assigned to Site 2 are identified by dashed lines. 
These assignments are those made by laser site selective fluorescence spectroscopy. Aside from Site 1 transitions, atmospheric water features are also observable for $^{5}$I$_{7}$.
Using the higher resolution afforded by the FTIR (0.075 cm$^{-1}$), a ground state splitting of 4.9 cm$^{-1}$ can be inferred. Superimposed on this ground state splitting are comparable splittings observed in the excited state. The sharpest absorption features are always observed for the lowest lying levels of a given multiplet since the higher levels broaden through phonon assisted relaxation. Thus, the lowest levels tend to exhibit the most clearly defined hyperfine splittings. Splittings of 2 cm$^{-1}$ are apparent for both $^{5}$I$_{7}$ (Y$_{1}$-Y$_{2}$) and $^{5}$I$_{6}$ (A$_{1}$-A$_{2}$). A larger splitting of 5 cm$^{-1}$ is measured for $^{5}$I$_{5}$ (B$_{1}$-B$_{2}$). Hyperfine structure cannot be observed on the scale presented in Figure~\ref{fig:absorption} but this will be presented and discussed below.

\clearpage

%%%%%%%%%%%%%%%%%%%%%%%%%%%%%%%%%%%%%%%%%%%%%%%%%%%%%%%
%%%%%Energy_levels_table
%%%%%%%%%%%%%%%%%%%%%%%%%%%%%%%%%%%%%%%%%%%%%%%%%%%%%%%

\begin{table*}[width=0.9\textwidth,cols=6,pos=h]
  \caption{\label{tab:laser_excitation}Energy level structure for Site 2 in Ho$^{3+}$:Y$_{2}$SiO$_{5}$. Energy levels are given in cm$^{-1}$. All the measurements were obtained at 10~K. Energy levels marked with an $\ast$ are tentative. The measurements have an uncertainty of 1 cm$^{-1}$.}
\footnotesize
\begin{tabular*}{\tblwidth}{@{} LLLLLL@{} }
  \toprule
    Multiplet & State & Energy & Multiplet & State & Energy \\
  \midrule
$^5$I$_8$	&	Z$_{1}$	&	0	&	$^5$I$_5$	&	B$_{1}$	&	11213	\\
	&	Z$_{2}$	&	5	&		&	B$_{2}$	&	11218	\\
	&	Z$_{3}$	&	39	&		&	B$_{3}$	&	11246	\\
	&	Z$_{4}$	&	46	&		&	B$_{4}$	&	11254	\\
	&	Z$_{5}$	&	78$^{*}$	&		&	B$_{5}$	&	11269	\\
	&	Z$_{6}$	&	85	&		&	B$_{6}$	&	11288	\\
	&	Z$_{7}$	&	180	&		&	B$_{7}$	&	-	\\
	&	Z$_{8}$	&	-	&		&	B$_{8}$	&	11308	\\
	&	Z$_{9}$	&	214	&		&	B$_{9}$	&	11312	\\
	&	Z$_{10}$	&	238	&		&	B$_{10}$	&	-	\\
	&	Z$_{11}$	&	279	&		&	B$_{11}$	&	-	\\
	&	Z$_{12}$	&	312$^{*}$	&		&		&		\\
	&	Z$_{13}$	&	323	&	$^5$F$_5$	&	D$_{1}$	&	15363	\\
	&	Z$_{14}$	&	353$^{*}$	&		&	D$_{2}$	&	15368	\\
	&	Z$_{15}$	&	359	&		&	D$_{3}$	&	15383	\\
	&	Z$_{16}$	&	383$^{*}$	&		&		&		\\
	&	Z$_{17}$	&	390	&		$^5$S$_2$,$^5$F$_4$	&	E$_{1}$	&	18408	\\
	&		&		&		&	E$_{2}$	&	18414		\\
$^5$I$_7$	&	Y$_{1}$	&	5135	&		&	E$_{3}$	&	18429		\\
	&	Y$_{2}$	&	5137	&		&		&		\\
	&	Y$_{3}$	&	5156	&		&		&		\\
	&	Y$_{4}$	&	-      	&		&		&		\\
	&	Y$_{5}$	&	5175	&		&		&		\\
	&	Y$_{6}$	&	5185	&		&		&		\\
	&	Y$_{7}$	&	5232	&		&		&		\\
	&	Y$_{8}$	&	5245	&		&		&		\\
	&	Y$_{9}$	&	5278	&		&		&		\\
	&	Y$_{10}$&	-	    &		&		&		\\
	&	Y$_{11}$&	-	    &		&	   &		\\
	&	Y$_{12}$&	5314	&		&		&		\\
	&	Y$_{13}$&	-	&		&		&		\\
	&	Y$_{14}$&	5398	&		&		&		\\
	&	Y$_{15}$&	5398	&		&		&		\\
	&		&		&		&		&		\\
$^5$I$_6$	&	A$_{1}$	&	8649	&		&		&		\\
	&	A$_{2}$	&	8651	&		&		&		\\
	&	A$_{3}$	&	8667	&		&		&		\\
	&	A$_{4}$	&	8695	&		&		&		\\
	&	A$_{5}$	&	8715	&		&		&		\\
	&	A$_{6}$	&	8736	&		&		&		\\
	&	A$_{7}$	&	8750	&		&		&		\\
	&	A$_{8}$	&	8771	&		&		&		\\
	&	A$_{9}$	&	8776	&		&		&		\\
	&	A$_{10}$	&	-	&		&		&		\\
	&	A$_{11}$	&	-	&		&		&		\\
	&	A$_{12}$	&	8886	&		&		&		\\
	&	A$_{13}$	&	8886	&		&		&		\\
	&		&		&		&		&		\\

  \bottomrule
  \end{tabular*}
\end{table*}

\clearpage

%%%%%%%%Emission%%%%%%%%%%%%%%%%%
%%%%%%%%%%%%%%%%%%%%%%%%%

%%%%%%%% Absorption%%%%%%%%%%%%%%%%%
%%%%%%%%%%%%%%%%%%%%%%%%%
\clearpage

%%%%%%%%%%%%%%%%%%%%%%%%%%%%%%%%%%%%%%%%%%%%
%%%%%%%%%Zeeman absorption spectroscopy
%%%%%%%%%%%%%%%%%%%%%%%%%%%%%%%%%%%%%%%%%%%%%
\subsection{Modelling of Zeeman absorption spectroscopy}
\label{sec:Zeeman_CF_analysis}

\begin{figure}
\centering
\includegraphics[width=0.48\textwidth]{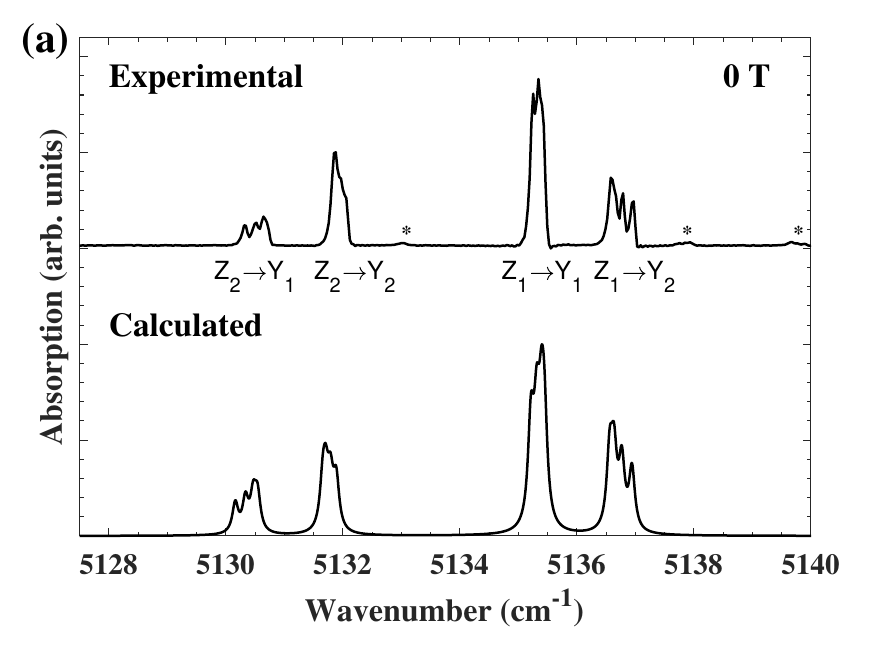}\\
\includegraphics[width=0.48\textwidth]{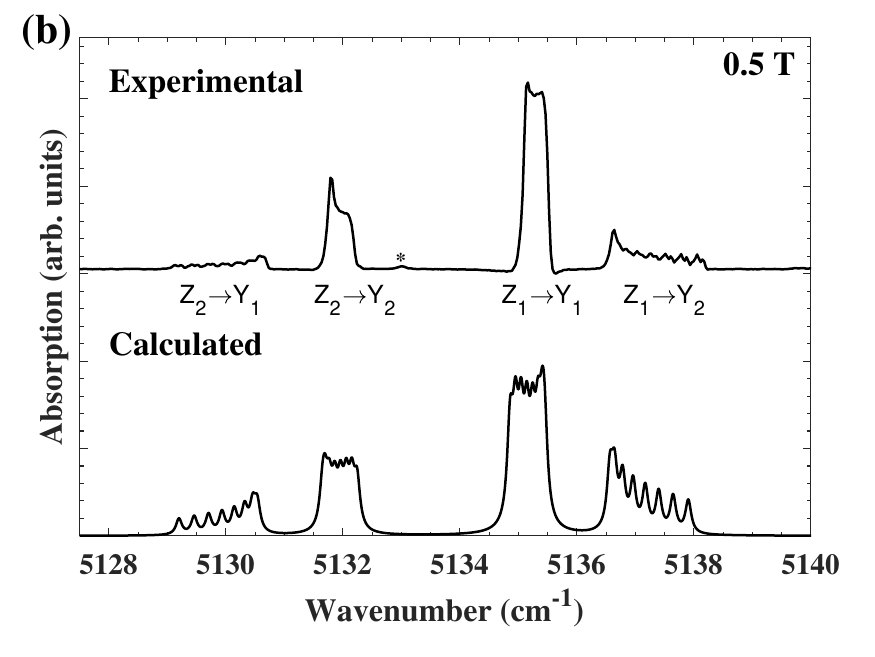}\\
\includegraphics[width=0.47\textwidth]{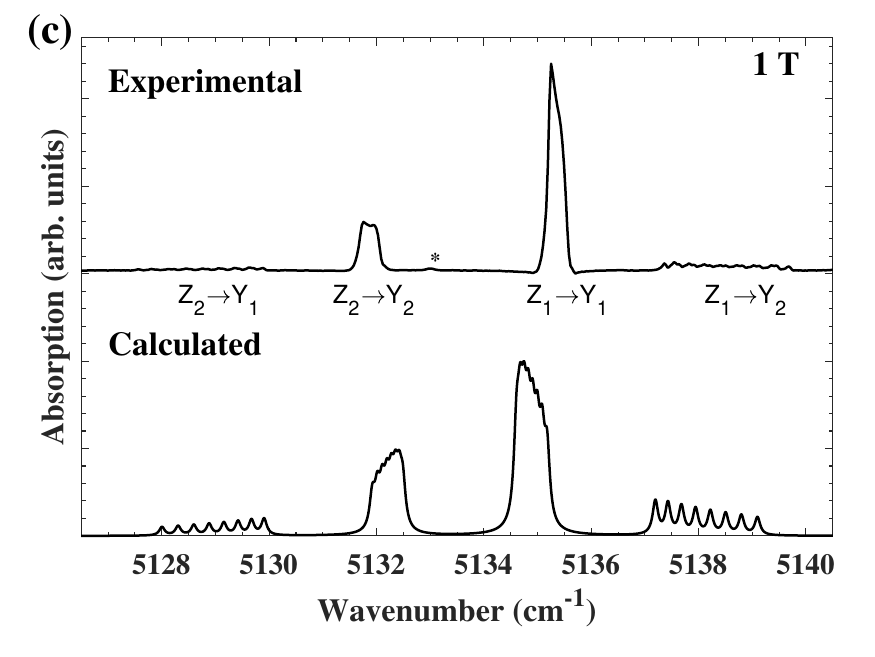}\\
[-5mm]
\vspace{4mm}
\caption{\label{fig:ZY_D1_Zeeman} 4.2~K Zeeman absorption spectrum of 
 $^5$I$_8$ (Z$_1$-Z$_2$) $\longrightarrow$$^5$I$_7$ (Y$_1$-Y$_2$) transitions for Site 2 of Ho$^{3+}$:Y$_{2}$SiO$_{5}$, (a) zero-field spectrum, (b) 0.5~T field spectrum, and (c) 1~T field spectrum with a magnetic field applied along the $D_2$ axis.} 
\end{figure}

\begin{figure}
\centering
\includegraphics[width=0.48\textwidth]{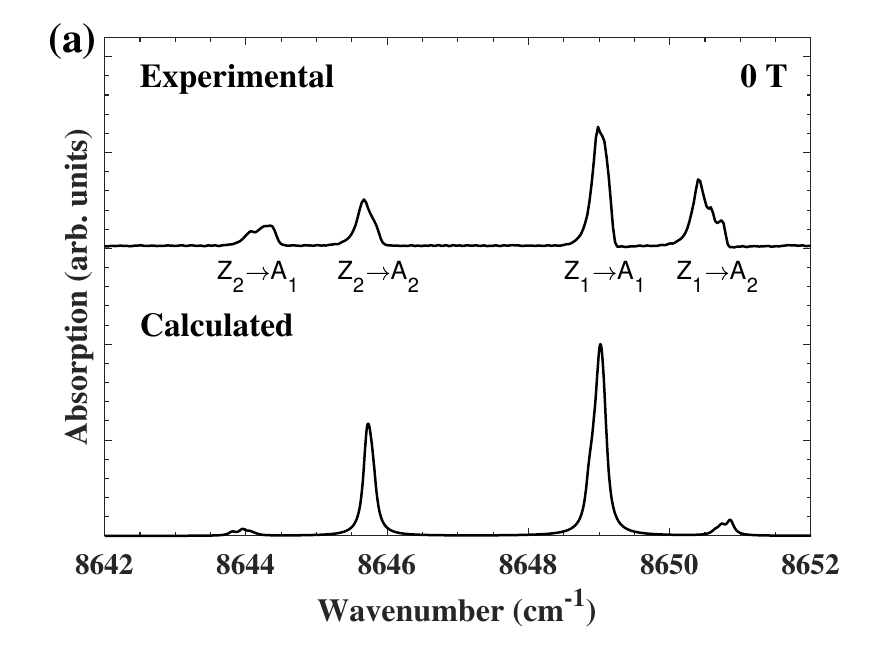}\\
\includegraphics[width=0.48\textwidth]{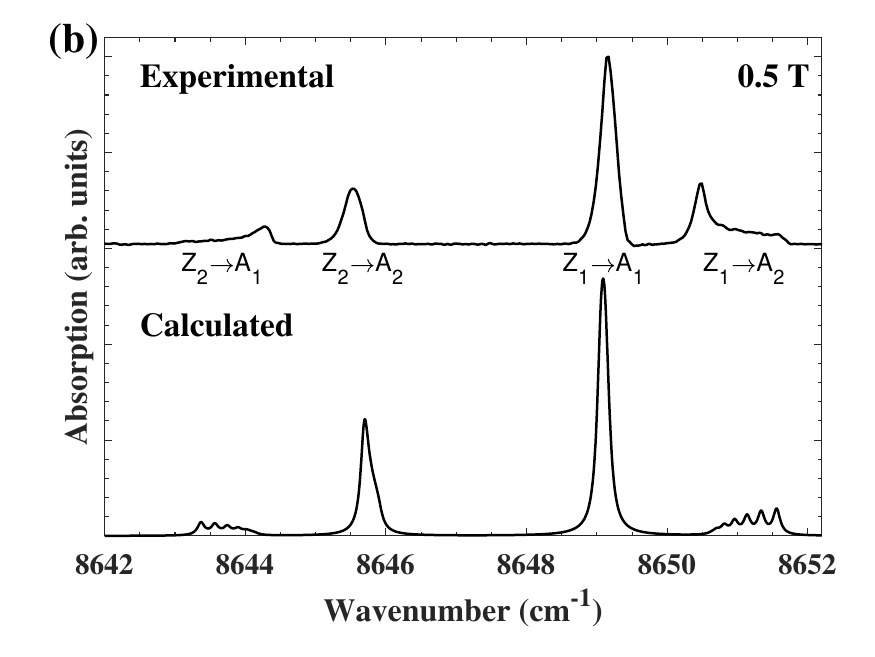}\\
\includegraphics[width=0.48\textwidth]{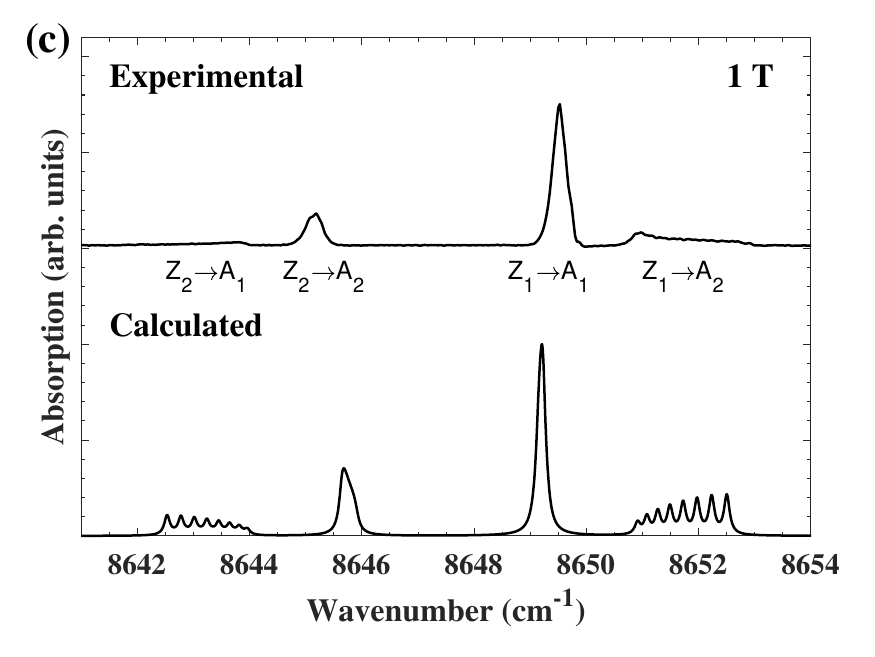}\\
[-5mm]
\vspace{4mm}
\caption{\label{fig:ZA_D1_Zeeman} 4.2~K Zeeman absorption spectrum of 
 $^5$I$_8$ (Z$_1$-Z$_2$) $\longrightarrow$$^5$I$_6$ (A$_1$-A$_2$) transitions for Site 2 of Ho$^{3+}$:Y$_{2}$SiO$_{5}$, (a) zero-field spectrum, (b) 0.5~T field spectrum, and (c) 1~T field spectrum with a magnetic field applied along the $D_2$ axis.} 
\end{figure}

\begin{figure*}
\centering
\includegraphics[width=\textwidth]{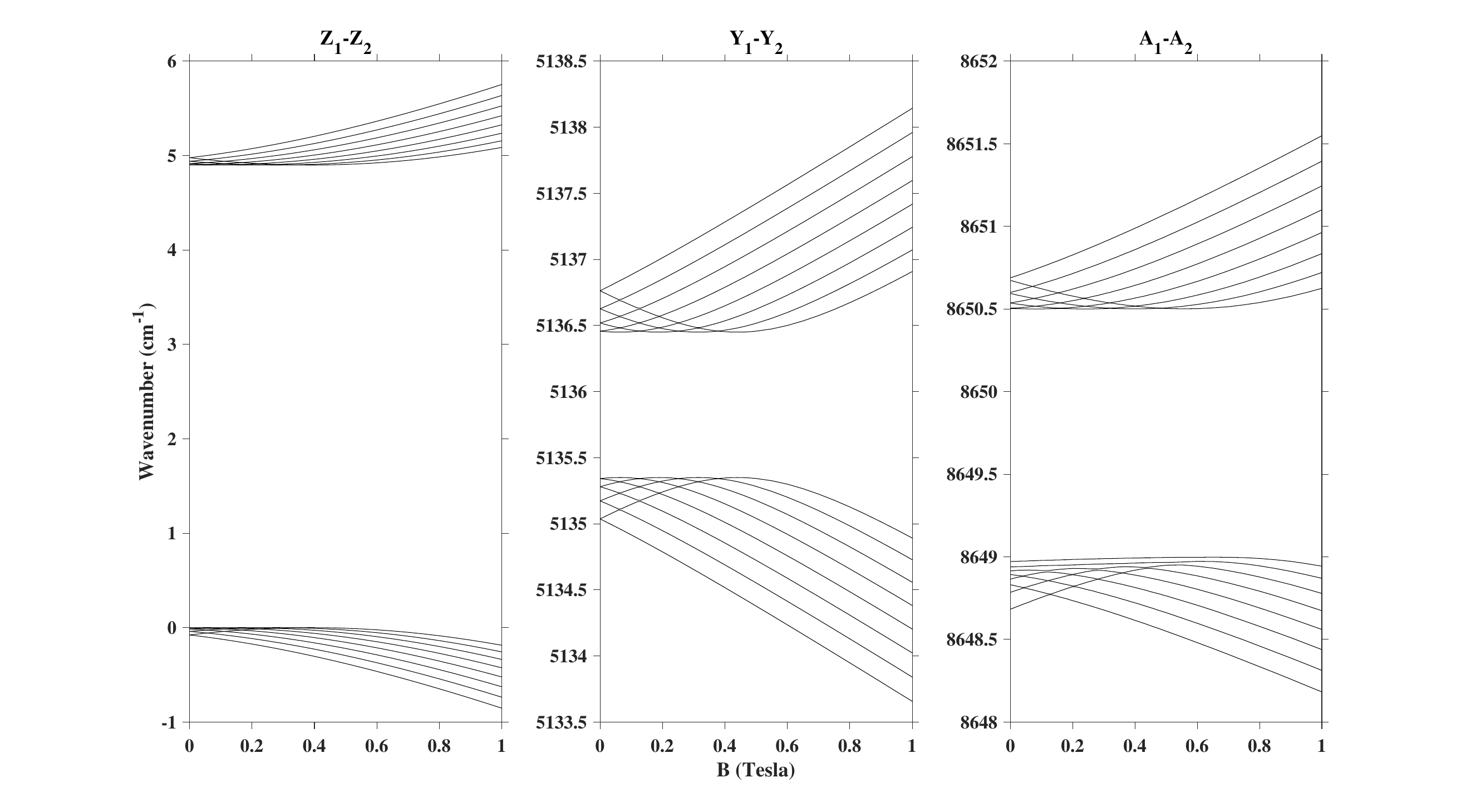}
\\
[-5mm]
\vspace{4mm}
\caption{\label{fig:D1_Z12A12profiles}Calculated Zeeman-hyperfine energy level structure of
  $^5$I$_8$ (Z$_1$-Z$_2$),  $^5$I$_7$ (Y$_1$-Y$_2$) and $^5$I$_6$ (A$_1$-A$_2$)
  for Site 2 of Ho$^{3+}$:Y$_{2}$SiO$_{5}$ with a magnetic field applied along the  $D_2$ axis.} 
\end{figure*}

Hyperfine structure was observed in absorption for both zero-field and for magnetic fields up to four Tesla. In this section we present data and
%{\color{red}
  simulated spectra.
%}
%{\color{red}[Text deleted.]}
%In the following section we discuss how these approximations break down at high magnetic fields. 

Figures \ref{fig:ZY_D1_Zeeman} and \ref{fig:ZA_D1_Zeeman} show the $^5$I$_8$ (Z$_1$-Z$_2$) $\longrightarrow$$^5$I$_7$ (Y$_1$-Y$_2$) and $^5$I$_8$ (Z$_1$-Z$_2$) $\longrightarrow$$^5$I$_6$ (A$_1$-A$_2$) transitions of Site 2 in Ho$^{3+}$:Y$_{2}$SiO$_{5}$ respectively, for magnetic fields of 0~T, 0.5~T, and 1~T directed along the $D_2$ axis. 
Figure~\ref{fig:D1_Z12A12profiles} shows the calculated energy levels of the states involved in the transitions and how they behave under the influence of a magnetic field. The closely spaced electronic singlets in both the ground and exited multiplets yield significant hyperfine splittings through the pseudo-quadrupole interaction~\cite{SagarMothkuri2021}. The splitting is partially resolved in the zero field spectra. The application of a magnetic field expands the hyperfine pattern for some transitions whilst collapsing the pattern on others. The simulated spectra in Figure ~\ref{fig:ZY_D1_Zeeman} and ~\ref{fig:ZA_D1_Zeeman} are in excellent agreement with the experimental data. Not only the eneries of the transitions, but also the relative intensities for the hyperfine splittings of, for example,  Z$_1 \rightarrow$Y$_2$ are well reproduced. The variations in intensity for different hyperfine components are caused by the mixing of wavefunctions through the pseudo-quadrupole interaction. The transitions in Figure \ref{fig:ZA_D1_Zeeman} are electric dipole, so while the energies are accurately predicted, the intensities are not. 

In high symmetries, optical transitions do not affect the nuclear spin, so only transitions with $\Delta{M_I}=0$ are allowed, resulting in a limited number of transitions between hyperfine levels \cite{Wells2004,Boldyrev2019}.  
In the low symmetry sites of Ho$^{3+}$ in Y$_{2}$SiO$_{5}$ the ${M_I}$ are mixed, and there is no selection rule restricting the number of allowed transitions.
However, both our experimental and calculated spectra show only four strong transitions at zero field. This approximate selection rule is reproduced in the model calculations. The eigenvectors of the hyperfine Hamiltonian for the ground and excited states are sufficiently similar that only four transitions with distinct energies at zero field, and 8 at non-zero fields, are calculated to have significant intensity.

\begin{figure*}
\centering
\includegraphics[width=0.495\textwidth]{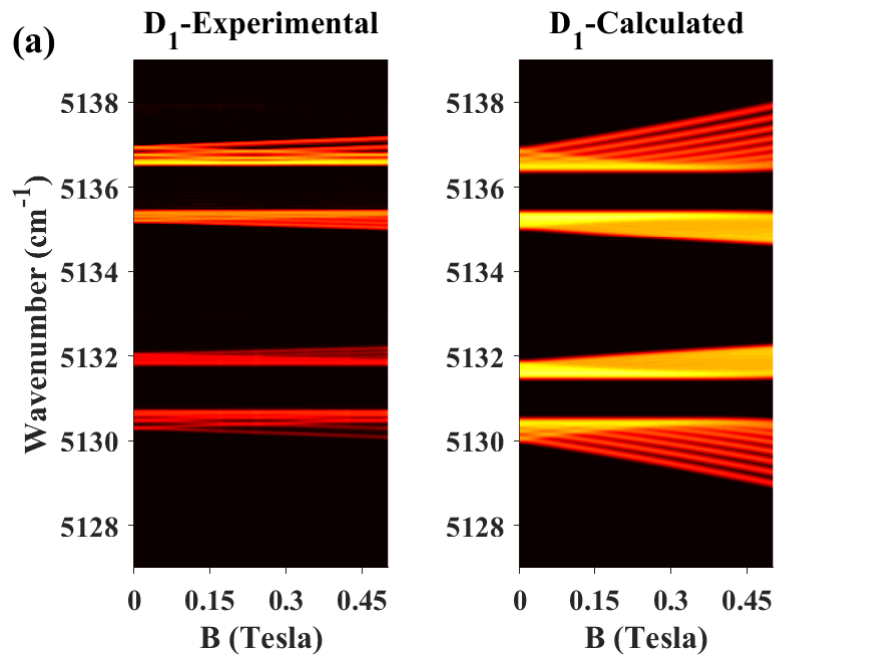}
\includegraphics[width=0.495\textwidth]{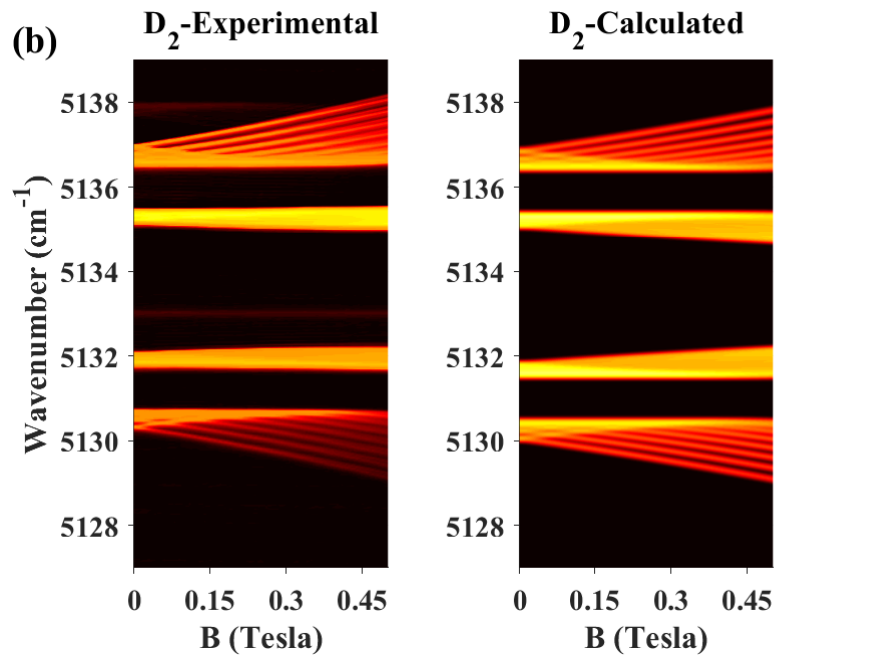}\\
\includegraphics[width=0.495\textwidth]{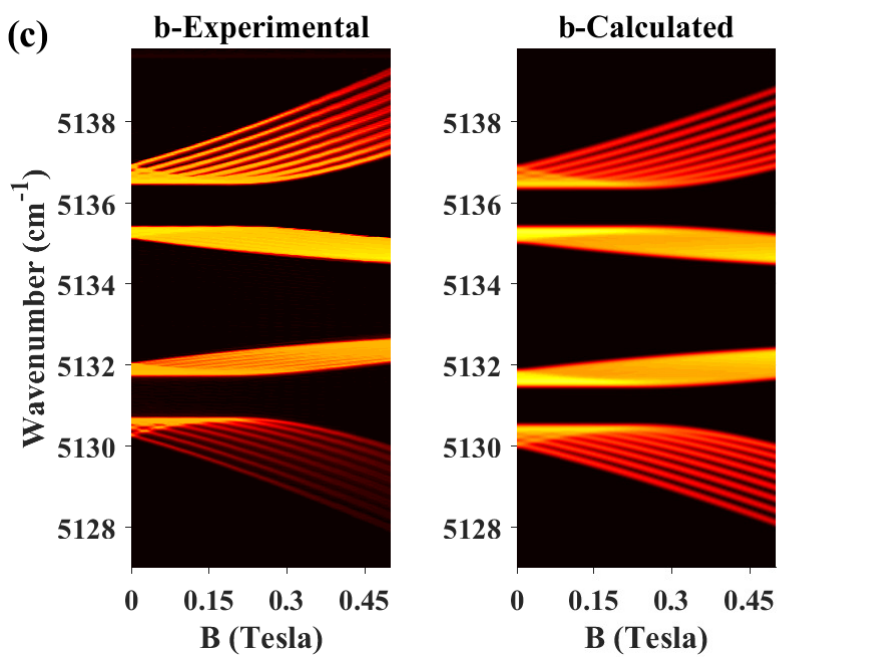}
\\
[-5mm]
\vspace{4mm}
\caption{\label{fig:ZY_Hotplot} 4.2~K experimental and calculated Zeeman-hyperfine splittings for the Z$_1$-Z$_2$$\longrightarrow$Y$_1$-Y$_2$ transitions of Site 2 for Ho$^{3+}$:Y$_{2}$SiO$_{5}$ with a magnetic field applied along (a) $D_1$, (b) $D_2$ and (c) $b$ axes.} 

\end{figure*}

\begin{figure*}
\centering
\includegraphics[width=0.495\textwidth]{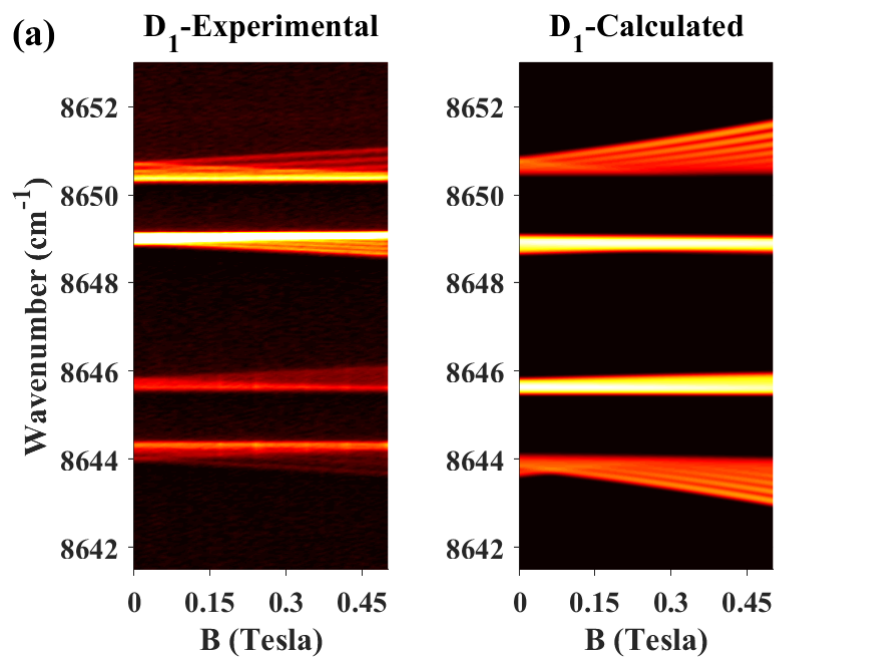}
\includegraphics[width=0.495\textwidth]{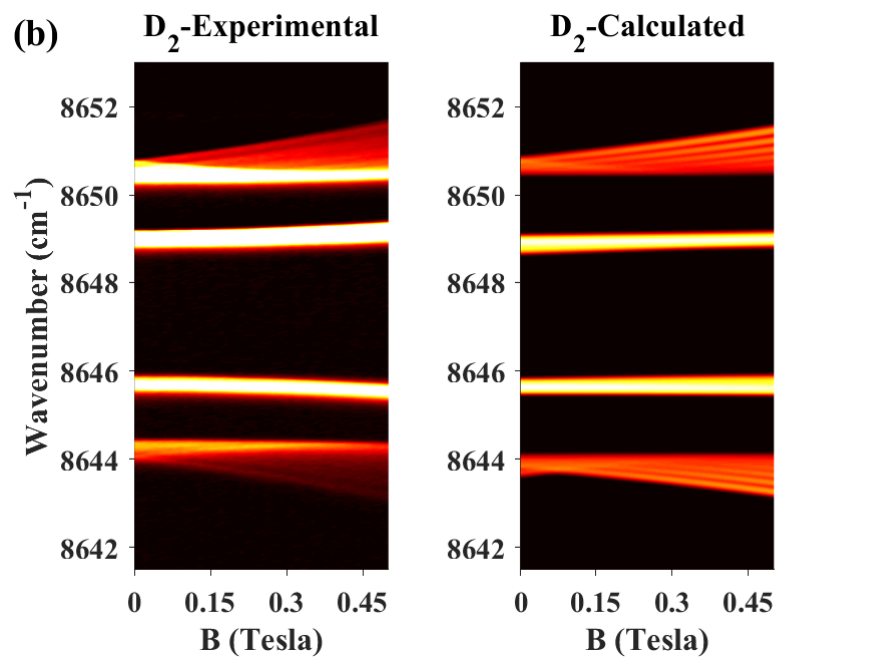}\\
\includegraphics[width=0.495\textwidth]{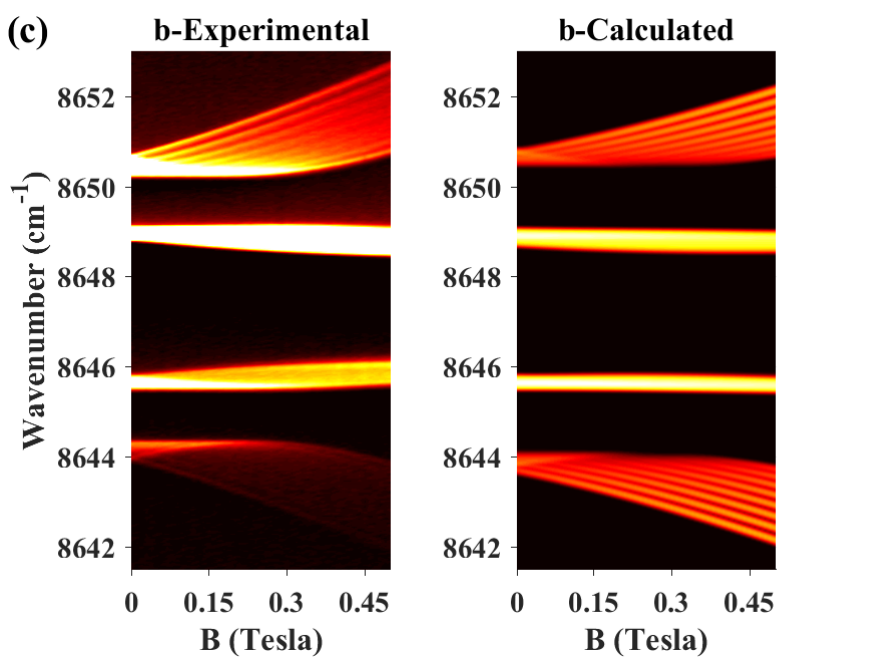}
\\
[-5mm]
\vspace{4mm}
\caption{\label{fig:ZA_Hotplot} 4.2~K experimental and calculated Zeeman-hyperfine splittings for the Z$_1$-Z$_2$$\longrightarrow$A$_1$-A$_2$ transitions of Site 2 for Ho$^{3+}$:Y$_{2}$SiO$_{5}$ with a magnetic field applied along (a) $D_1$, (b) $D_2$ and (c) $b$ axes.} 

\end{figure*}

Figure~\ref{fig:ZY_Hotplot} shows the experimental and calculated spectra for the $^5$I$_8$ (Z$_1$-Z$_2$) $\longrightarrow$$^5$I$_7$ (Y$_1$-Y$_2$) transitions of Site 2 in Ho$^{3+}$:Y$_{2}$SiO$_{5}$ with a magnetic field applied along $D_1$, $D_2$ and $b$ axes.  Similarly, Figure~\ref{fig:ZA_Hotplot} shows the experimental and calculated spectra for the $^5$I$_8$ (Z$_1$-Z$_2$) $\longrightarrow$$^5$I$_6$ (A$_1$-A$_2$) transitions of Site 2 in Ho$^{3+}$:Y$_{2}$SiO$_{5}$ with a magnetic field applied along $D_1$, $D_2$ and $b$ axes. The magnetic field is varied in steps of 0.01~T. The calculated spectra reproduce the experimental spectra quite well. The magnitude of the matrix elements of the angular-momentum operators is reflected in the  magnetic-hyperfine splittings. Matrix elements of ${J}_{x}$ and  ${J}_{y}$ (along $D_1$ and $D_2$) respectively, are small compared to matrix elements of  ${J}_{z}$ (along $b$). This is reflected in the experimental and calculated spectra in Fig.~\ref{fig:ZY_Hotplot} and \ref{fig:ZA_Hotplot}.

\clearpage

\section{Conclusions}
Using a combination of laser site selective excitation and absorption spectroscopy, a detailed analysis of the energy level structure for Site 2 in Ho$^{3+}$ doped Y$_{2}$SiO$_{5}$ has been performed. Due to the presence of ground and excited psuedo-doublets, hyperfine patterns are observed in high resolution absorption spectra which appear to obey an approximate selection rule, despite the low symmetry.
%{\color{red}[Text deleted.]}
% This approximate selection rule breaks down with the application of a magnetic field. 

\section*{Acknowledgments}
SM acknowledges support from the University of Canterbury and Dodd-Walls Centre for Photonic and Quantum Technologies in the form of a PhD Scholarship. The technical assistance of Mr Graeme MacDonald (UoC, NZ), Mr Stephen Hemmingson, and Mr Robert Thirkettle is gratefully acknowledged.
 
%\bibliographystyle{elsarticle-num}
%\bibliography{hoyso}

\end{document}